\documentclass[journal]{IEEEtran}
\ifCLASSINFOpdf
  \usepackage[pdftex]{graphicx}
\else
\fi
\usepackage[cmex10]{amsmath}
\usepackage{amssymb}
\usepackage{mathtools}
\usepackage{array}

\usepackage{algpseudocode,algorithm,algorithmicx}
\usepackage{float}
\usepackage{tabularx}
\usepackage{mathrsfs} 
\usepackage{psfrag}
\usepackage{tikz}
\usetikzlibrary{chains, arrows,shapes,positioning,arrows,decorations.markings,calc}
\usepackage{pgfplots}
\pgfplotsset{compat=newest}
\usepgfplotslibrary{groupplots}
\usepgfplotslibrary{polar}
\usepackage[pdfusetitle,hidelinks, pdfauthor={Kilian Roth}, pdfsubject={Arbitrary Beam Synthesis of Hybrid Beamforming Systems for Beam Training}, breaklinks=true]{hyperref}
\usepackage{gensymb}
\usepackage{xcolor}
\usepackage[shortcuts, acronym]{glossaries}
\usepackage[caption=false]{subfig}

\begin{document}
%
\title{Arbitrary Beam Synthesis of Hybrid Beamforming Systems for Beam Training}
\author{Kilian~Roth,~\IEEEmembership{Member,~IEEE,}
        Josef~A.~Nossek,~\IEEEmembership{Life Fellow,~IEEE}%
\thanks{K. Roth is with Next Generation and Standards, Intel Deutschland GmbH, Neubiberg 85579, Germany (email: $\{$kilian.roth$\}$@intel.com)}%
\thanks{K. Roth and J. A. Nossek are with the Department of Electrical and Computer Engineering, Technical University Munich, Munich 80290, Germany (email: $\{$kilian.roth, josef.a.nossek$\}$@tum.de)}%
\thanks{J. A. Nossek is with Department of Teleinformatics Engineering, Federal University of Ceara, Fortaleza, Brazil}}%
\maketitle
\newacronym{ADC}{ADC}{Analog-to-Digital-Converter}
\newacronym{A/D}{A/D}{Analog/Digital}
\newacronym{OFDM}{OFDM}{Orthogonal Frequency Domain Multiplexing}
\newacronym{MSE}{MSE}{Mean Square Error}
\newacronym[plural=UEs,firstplural=User Equipments (UEs)]{UE}{UE}{User Equipment}
\newacronym{RF}{RF}{Radio Frequency}
\newacronym{RFE}{RFE}{Radio Front-End}
\newacronym{LTE}{LTE}{Long Term Evolution}
\newacronym{MMSE}{MMSE}{Minimum Mean Square Error}
\newacronym{PDP}{PDP}{Power Delay Profile}
\newacronym{SNR}{SNR}{Signal to Noise Ratio}
\newacronym{MU-MIMO}{MU-MIMO}{Multi User - Multiple Input Multiple Output}
\newacronym{CIR}{CIR}{Channel Impulse Response}
\newacronym{PA}{PA}{Power Amplifier}
\newacronym{NLP}{NLP}{NonLinear Programing}
\newacronym{MINLP}{MINLP}{Mixed Integer Non-Linear Programing}
\newacronym{ULA}{ULA}{Uniform Linear Array}
\newacronym{mmWave}{mmWave}{millimeter Wave}
\newacronym{MU MIMO}{MU MIMO}{MultiUser MIMO}
\begin{abstract}
For future \ac*{mmWave} mobile communication systems, the use of analog/hybrid beamforming is envisioned to be an important aspect.
The synthesis of beams is a key technology to enable the best possible operation during beam search, data transmission and
\ac*{MU MIMO} operation. The method for synthesizing beams developed in this work is based on previous work in radar technology considering only phase array antennas.
With this technique, it is possible to generate a desired beam of any shape with the constraints of the desired target transceiver antenna frontend.
It is not constraint to a certain antenna array geometry, and can handle 1D, 2D and even 3D antenna array geometries, e.g. cylindrical arrays. 
The numerical examples show that the method can synthesize beams by considering 
a user defined trade-off between gain, transition width and passband ripples.
\end{abstract}
%
%
%
%
%
\section{Introduction}
To satisfy the ever increasing data rate demand, the use of the available bandwidth in the 
\ac*{mmWave} frequency range is considered to be an essential part of the next generation mobile broadband standard \cite{FIVEDISRUPTIVE}. 
To attain a similar link budget, the effective antenna aperture of a \ac*{mmWave} system must be comparable to current
systems operating at a lower carrier frequency. 
Since the antenna gain, and thus the directivity 
increases with the aperture, an antenna array is the only solution to achieve a high effective aperture, while maintaining a $360^\circ$ coverage.

The antenna array combined with the large bandwidth is a big challenge for the hardware implementation as
the power consumption limits the design space. Analog or hybrid beamforming are considered to be
possible solutions to reduce the power consumption. These solutions are based on the concept of phased array antennas. 
In this type of systems the signal of multiple antennas are phase shifted, combined and afterwards converted into the analog baseband followed by an A/D conversion. 
If the signals are converted to only one digital signal we speak of analog beamforming, otherwise hybrid beamforming is used.
For the transmission the digital signal is converted to a analog baseband signal, followed by a up-conversion. Afterwards,
the signal is split into multiple signals, separately phase shifted, ampflied and then transmitted at the antennas. 

To utilize the full potential of the system, it is essential that the beams of Tx and Rx are aligned.
Therefore, a trial and error procedure is used to align the beams of Tx and Rx \cite{WIGIGSTDORIGINAL, 80211ayBF}.
This beam search procedure does either utilize beams of different width with additional feedback or many beams of the same width with 
only one feedback stage \cite{Palacios2016}. In both cases the beams with specific width, maximum gain and flatness need to be designed.

Based on requirements on the beam shape, this work formulates an optimization problem similar to \cite{Scholnik2016, Morabito2012}. 
Afterwards the optimization problem is solved numerically. 
This work includes the specific constraints of hybrid beamforming and low resolution phase shifters. 
In \cite{Palacios2016}, the authors approximate a digital beamforming vector by a hybrid one. We generate our beam by approximating a desired beam instead.

The superscript $s$ and $f$ are used to distinguish between sub-array and fully-connected hybrid beamforming.  
Bold small $\boldsymbol{a}$ and capital letters $\boldsymbol{A}$ are used to represent vectors and matrices. The notation 
$[\boldsymbol{a}]_n$ is the $n$th element of the vector $\boldsymbol{a}$. 
The superscript $T$ and $H$ represent the transpose and hermitian operators. 
The symbol $\circ$ is the Hadamard product.
\begin{figure*}[!t]
\centering
\normalsize
\input{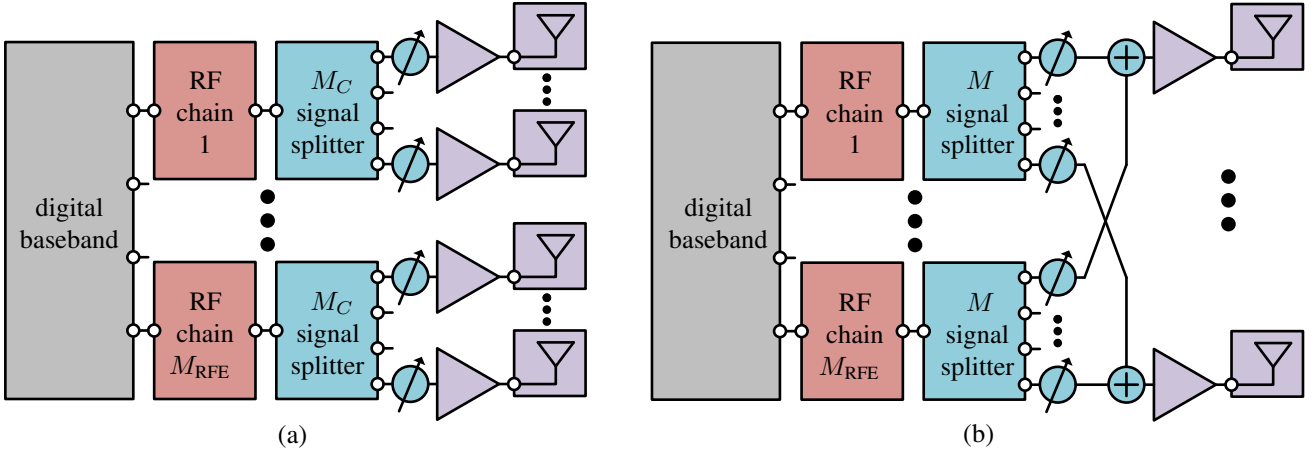}
\vspace*{-0.4cm}
	\caption{System model of hybrid beamforming transmitter with $M$ antennas and $M_{\text{RFE}}$ RF-chains for the sub-array (a) and the fully-connected (b) case.}
	\label{fig:SystemModel}
\hrulefill
\vspace*{-0.5cm}
\end{figure*}
\section{Optimum Beam Synthesis}
In the following we will develop a strategy to synthesis arbitrary beams based on the formulation an optimization problem.
Furthermore, we show how different constraints can be used to model the restrictions of different systems.
\subsection{Objective function}
The array factor $A(\boldsymbol{u}, \boldsymbol{a})$ of an antenna array is defined as
\begin{equation}
	A(\boldsymbol{u}, \boldsymbol{a}) = \boldsymbol{a}^T \boldsymbol{p}(\boldsymbol{u})~,~\left[\boldsymbol{p}(\boldsymbol{u})\right]_n = e^{j\frac{2\pi}{\lambda}x_n(\boldsymbol{u})},
\end{equation}
where $\boldsymbol{a}$ is the beamforming vector, $\boldsymbol{u}$ is the spatial direction combining the azimuth and elevation angle. 
The scalar $x_n(\boldsymbol{u})$ 
is the distance from the location of antenna element $n$ to the plane defined by the normal vector $\boldsymbol{u}$ and a reference point. A common choice for the reference point is the position of the first
antenna, in this case $x_1(\boldsymbol{u}) = 0$.

The objective of synthesizing an arbitrary beam pattern can be formulated as a weighted $L^p$ norm
between the desired pattern $D(\boldsymbol{u})$ and the absolute value of the actual array factor $\vert A(\boldsymbol{u}, \boldsymbol{a})\vert$
\begin{equation}
	f(\boldsymbol{a}) = \left( \int W^p(\boldsymbol{u}) \left\vert\left\vert A(\boldsymbol{u}, \boldsymbol{a})\right\vert - D(\boldsymbol{u})\right\vert^p d\boldsymbol{u}\right)^{\frac{1}{p}},
\end{equation}
where $W(\boldsymbol{u})$ is the weighting.
This objective function itself is convex over its domain, but the constraints on $\boldsymbol{a}$ shown in the following subsections lead to a non-convex optimization problem. 
This problem formulation ignores the phase of the array factor, since we require only the magnitude to be of a specific shape.

By only optimizing over the array factor we don't take the pattern of the antennas into account. 
As described in \cite{Scholnik2016} to account for an antenna pattern it is only necessary 
to divide $D(\boldsymbol{u})$ and $W(\boldsymbol{u})$ 
by the pattern of the antenna elements.
\subsection{Constraints}
We consider two different hybrid beamforming designs. These are the systems currently considered in literature \cite{Palacios2016, HBFMAG}. 
In the first case, all $M$ antennas are divided into groups of size $M_C$. Each subgroup consists of one \ac*{RF} chain, an $M_C$ signal 
splitter followed by a phase shifter and a \ac*{PA} at each antenna (see Figure \ref{fig:SystemModel} (a)). In total there are $M_{\text{RFE}}$ \ac*{RF} chains. This restricts the beamforming vector $\boldsymbol{a}$ to have the form
\begin{equation}
	\boldsymbol{a} = \boldsymbol{W}^s \boldsymbol{\alpha}^s = 
	\begin{bmatrix}
		\boldsymbol{w}^s_1  & \boldsymbol{0}   & \cdots  & \boldsymbol{0}  \\
		\boldsymbol{0} & \boldsymbol{w}^s_2  & \ddots  & \boldsymbol{0}  \\
		\vdots & \vdots  & \ddots &\vdots    \\
		\boldsymbol{0} & \cdots  & \boldsymbol{0} & \boldsymbol{w}^s_{M_{\text{RFE}}} \\
	\end{bmatrix}
	\begin{bmatrix}
		\alpha^s_1  \\ \alpha^s_2 \\ \vdots  \\ \alpha^s_{M_{\text{RFE}}}
	\end{bmatrix}, 
\end{equation}
where $\boldsymbol{\alpha}^s \in \mathbb{R}^{M_{\text{RFE}} \times 1}$ and the vectors $\boldsymbol{w}^s_i$ models the analog phase shifting of group $i$ and therefore has the form
\begin{equation}
	\boldsymbol{w}^s_i = 
	\begin{bmatrix}
		e^{j\theta^s_{1,i}} & e^{j\theta^s_{2,i}} & \cdots &  e^{j\theta^s_{{M_C},i}}
	\end{bmatrix}^T.
\end{equation}

In the second case, each of the \ac*{RF} chain is connected to an $M$ signal splitter followed by a phase shifter for each antenna
(see Figure \ref{fig:SystemModel} (b)). At each antenna, the phase shifted signal from each \ac*{RF} chain is combined and 
then amplified by a \ac*{PA} followed by the antenna transmission. 
With this system architecture the beamforming vector $\boldsymbol{a}$ can be decomposed into
\begin{equation}
	\begin{gathered}
	\boldsymbol{a} = \boldsymbol{W}^f \boldsymbol{\alpha}^f = 
	\begin{bmatrix}
		\boldsymbol{w}^f_1 & \boldsymbol{w}^f_2 & \cdots & \boldsymbol{w}^f_{M_{\text{RFE}}}
	\end{bmatrix}
	\boldsymbol{\alpha}^f
	\\ =
	\begin{bmatrix}
		e^{j\theta^f_{1, 1}} & e^{j\theta^f_{1, 2}} & \cdots &  e^{j\theta^f_{1, M_{\text{RFE}}}} \\
		e^{j\theta^f_{2, 1}} & e^{j\theta^f_{2, 2}} & \cdots &  e^{j\theta^f_{2, M_{\text{RFE}}}} \\
		\vdots & \vdots  & \ddots &\vdots    \\
		e^{j\theta^f_{M, 1}} & e^{j\theta^f_{M, 2}} & \cdots &  e^{j\theta^f_{M, M_{\text{RFE}}}} \\
	\end{bmatrix}
	\begin{bmatrix}
		\alpha^f_1  \\ \alpha^f_2 \\ \vdots  \\ \alpha^f_{M_{\text{RFE}}}
	\end{bmatrix}
	\end{gathered},
\end{equation}
with $\boldsymbol{\alpha}^f \in \mathbb{R}^{M_{\text{RFE}} \times 1}$.

To limit the maximum output power of the \ac*{PA}s, we need to include the following constraints
\begin{equation}
	[\boldsymbol{a}]_m \leq 1 ~ \forall m = \{1, 2, \cdots, M\}.
\end{equation}
It is important to keep in mind that this restriction is after the hybrid beamforming, therefore, it is a 
nonlinear constraint restricting output-power of the \ac*{PA}.
Another way to bound the output power is a sum power constraint of the form
\begin{equation}
	\vert \vert \boldsymbol{a} \vert\vert^2 \leq 1.
\end{equation}

It is also possible that the resolution of the phase shifters is limited. This means that the values of $\theta^s_{i,j}$ are from a
finite set of possibilities
\begin{equation}
	\theta^s_{i,j} = -\pi + k_{i,j}\frac{2\pi}{K}~~\forall i,j~\text{and}~k_{i,j} \in \{0, 1, \cdots, K-1\},
\end{equation}
where $K$ is the number of possible phases. 
A possible phase shift in the digital domain needs to be taken into account. In the case
without quantization, this phase shift is redundant with the analog phase shift. 
Therefore, in addition to the scaling $\boldsymbol{\alpha}^f$ or $\boldsymbol{\alpha}^s$, we need to 
take a phase shift $\boldsymbol{\xi}^f$ or $\boldsymbol{\xi}^s$ into account. 
For the case of sub-array hybrid beamforming with limited resolution \ac*{RF} phase shifters the beamforming vector $\boldsymbol{a}$ takes the form
\begin{equation}
	\boldsymbol{a} = \boldsymbol{W}^s\left(\boldsymbol{\alpha}^s \circ \boldsymbol{\xi}^s\right),
\end{equation}
where $\boldsymbol{\xi}^s$ are the digital phase shifts defined as
\begin{equation}
	\boldsymbol{\xi}^s = [e^{j\xi_1^s}, e^{j\xi_2^s}, \cdots, e^{j\xi^s_{M_{\text{RFE}}}}]^T.
\end{equation}
The formulation for the fully-connected case does also contain addition phase shifts in the digital baseband signals.
\subsection{Problem Formulation}
\begin{figure}
	\centering
	\includegraphics{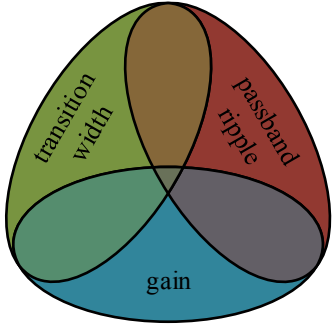}
	\caption{Illustration of the trade-off associated with the beam pattern synthesis.}
	\label{fig:tradeoff}
\end{figure}
Combining the objective function with the constraints associated with the hardware capabilities lead to the following optimization problem
\begin{equation}
	\begin{array}{l} 
		\min f(\boldsymbol{a}) \\
		\text{s.t.}~\boldsymbol{g}(\boldsymbol{a}) \leq \boldsymbol{0} ~,~\boldsymbol{h}(\boldsymbol{a}) = \boldsymbol{0},
	\end{array}
\end{equation}
where $\boldsymbol{g}(\boldsymbol{a})$ and $\boldsymbol{h}(\boldsymbol{a})$ are the constraints modelling the desired hardware capabilities.
It is important to mention that beam synthesis is a similar procedure as digital filter design, therefore we us the terminology of digital filter design.
The weighting $W(\boldsymbol{u})$, the desired pattern $D(\boldsymbol{u})$ and the choice of $p$ in $f(\boldsymbol{a})$, determine
which point in the trade-off gain, passband ripple and transition width is going to be targeted as shown in Fig. \ref{fig:tradeoff}.
\section{Numerical results}
To compare the designed beams we need to first define some metrics to quantify the difference between them.
Some of these metrics are similar to the ones defined in \cite{Donno2016}. 
The first one is the \textit{average gain} in the desired direction. Directly connected to the average gain is the \textit{maximum ripple} of the array factor in the desired directions.
For more reliable results, the transition region is excluded from the search of the maximum ripple.
A very important criteria to evaluate the performance of a beam for initial access is the \textit{overlap of adjacent beams} of the same width. Here we evaluate the
area at which the distance between two beams is less than 5 dB relative to the total area of one beam. 
The last measure is the \textit{maximum side-lobe} relative to the average gain in the desired directions.

In the following, beams synthesized by the described method are shown.
For all systems, the transmitter is equipped with $M_{\text{RFE}} = 4$ \ac*{RF}-chains, connected to 64 Antenna elements, forming an \ac*{ULA} with half-wavelength inter-element spacing. 
Since the antenna array is one dimensional, it is sufficient to look at only one spatial direction. All plots refer to angle $\psi = \frac{\lambda}{2}\sin(\phi)$, where
$\phi$ is the geometric angle between a line connecting all antennas and the direction of a planar wavefront.

For each system, three beams of width $b = \pi, \pi/2, \pi/4$ are synthesized. But it is important to mention that the beams in Fig. \ref{FIG:BEAMPATTERNENQ} and
\ref{FIG:BEAMPATTERNFNQ} are not designed to be used simultaneously. In contrast, the beams in Fig. \ref{FIG:BEAMPATTERNFQ} and \ref{FIG:BEAMPATTERNFQIMDEA} can be simultaneously used. 
For an \ac*{ULA}, the spatial direction $\boldsymbol{u}$ is fully represented by $\psi$, therefore $W(\boldsymbol{u})$, $D(\boldsymbol{u})$ and $A(\boldsymbol{u}, \boldsymbol{a})$ depend only on $\psi$. 
Since the magnitude of each element of $\boldsymbol{a}$
is less or equal to one, if a perfect flat beam without sidelobes could be constructed, it would have the array-factor $D_{\text{max}} = \sqrt{N2\pi/b}$.
As also described in \cite{Scholnik2016}, such a beam cannot be realized, therefore $D(\psi)$ is equal to $\beta D_{\text{max}}$ at the desired directions and equal to zero, elsewhere. 
The parameter $\beta$ ensures the feasibility of a solution.
The weighting of different parts of the beam pattern $W(\psi)$ is uniformly set to 1, except for a small transition region enclosing the desired directions. 
For all systems, we set $p = 4$ in the objective function to ensure equal gain and side lobe ripples. The integral of the objective function over all spatial directions in the
objective function is approximated by a finite sum. To ensure a sufficient approximation, the interval is split into 512 elements. As described in \cite{Scholnik2016},
the computational complexity can be significantly reduced by reformulating the problem to use FFT/IFFTs to calculate $A(\psi, \boldsymbol{a})$ and the derivatives of the objective function.

For each system, the optimization process was started by considering several initializations. Since the used  \ac*{NLP} and \ac*{MINLP} solvers only guarantee to find
a local minimum for a non-convex problem, the results were compared and the implementation leading towards the minimum objective function was selected.
The metrics to compare the performance of different beams is shown in Table \ref{TAB:measuretab} alongside a reference to the respective Fig.. 

In Fig. \ref{FIG:BEAMPATTERNENQ} and \ref{FIG:BEAMPATTERNFNQ} the synthesized beams for sub-array and 
fully-connected hybrid beamforming are shown. For (a), (b) and (c) the gain penalty $\beta$ was selected to be 3 dB, 2 dB and 2 dB, respectively. 
Compared to the fully-connected case, sub-array hybrid
beamforming is characterized by more gain ripples and higher sidelobe energy, while having the same transition width. 

In Fig. \ref{FIG:BEAMPATTERNFQ} and \ref{FIG:BEAMPATTERNFQIMDEA} fully-connected hybrid beamforming with quantized phase shifters was applied.
The beams are designed with the method described in Fig. \ref{FIG:BEAMPATTERNFQIMDEA}.
The beam in both figures is optimized to simultaneously transmit us both shown beams at each stage (a), (b) and (c). The power constraint for this case is also different,
in this case only the sum power is constraint to be less or equal to 1. For our evaluation we used the same constraints.

In Fig. \ref{FIG:BEAMPATTERNFQIMDEA}, and, especially in (a) there are multiple points where both beams almost overlap. 
In these directions an estimation of the link quality achieved with both beams is going to be very similar. 
This can possibly lead to a wrong decision and, in its turn, to large errors in a multi-stage beam
training procedure. On the contrary, the solution evaluated in Fig. \ref{FIG:BEAMPATTERNFQ} offers a sharper transition. The
stop directions attenuation is also close to uniform to enable a predictable performance. The only disadvantage is the larger ripples inside the center main beam.

The shortcomings which are observed in Fig. \ref{FIG:BEAMPATTERNFQIMDEA} are introduced during the generation of $\boldsymbol{a}$. 
As described in \cite{Palacios2016} this method approximates a version of $\boldsymbol{a}_d$ generated with the assumption of full digital beamforming. 
Since for a low number of \ac*{RF}-chains this vector cannot be well approximated, the resulting beam pattern does not correspond well to the
desired one. It is also important to mention that there is no one-to-one mapping between the error in approximating $\boldsymbol{a}_d$ and
the errors of the corresponding beam. 
As shown in \cite{Palacios2016}, the method works well if $\boldsymbol{a}_d$ can be well approximated by a larger number of \ac*{RF} chains.
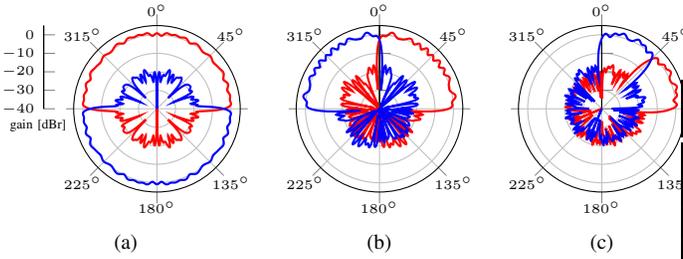
\begin{figure}
\vspace*{-0.5cm}
\centering
\subfloat[][]{
\begin{tikzpicture}
\begin{polaraxis}[
	scale only axis=true,
 	width = 0.25*\columnwidth,
	height = 0.25*\columnwidth,
  	rotate=-90,
	grid=both,
	xticklabel=$\pgfmathprintnumber{\tick}^{\circ}$,
   	xtick={0, 45, 135, 180, 225, 315},
	minor xtick={90, 270},
	x dir=reverse,
	xticklabel style={anchor=-\tick-90, font=\tiny},
	xtick style={font=\tiny},
   	ytick={-40,-30,-20,-10,0},
   	ymin=-40, ymax=5,
   	y coord trafo/.code=\pgfmathparse{#1+40},
   	y coord inv trafo/.code=\pgfmathparse{#1-40},
	ylabel style={font=\tiny, yshift=-0.11*\columnwidth, xshift=-0.20*\columnwidth},
	ylabel={gain [dBr]},
	yticklabel style={anchor=east, xshift=-0.17*\columnwidth, font=\tiny},
  	y axis line style={yshift=-0.17*\columnwidth,},
   	ytick style={yshift=-0.17*\columnwidth,font=\tiny},
]
\addplot [no markers, thick, red] table [x=y, y=x] {./SimulationResults/PlotData/MR_64_N_16_Stage_1_Beam_1_hbf.txt};
\addplot [no markers, thick, blue] table [x=y, y=x] {./SimulationResults/PlotData/MR_64_N_16_Stage_1_Beam_2_hbf.txt};
\end{polaraxis}
\end{tikzpicture}
}
\subfloat[][]{
\begin{tikzpicture}
\begin{polaraxis}[
	scale only axis=true,
 	width = 0.25*\columnwidth,
	height = 0.25*\columnwidth,
	grid=both,
   	rotate=-90,
	xticklabel=$\pgfmathprintnumber{\tick}^{\circ}$,
   	xtick={0, 45, 135, 180, 225, 315},
	minor xtick={90, 270},
	x dir=reverse,
   	xticklabel style={anchor=-\tick-90, font=\tiny},
	xtick style={font=\tiny},
   	ytick={-40,-30,-20,-10,0},
   	ymin=-40, ymax=5,
   	y coord trafo/.code=\pgfmathparse{#1+40},
   	y coord inv trafo/.code=\pgfmathparse{#1-40},
	yticklabels={,,},
	ymajorticks=true,
	yminorticks=true,
]
\addplot [no markers, thick, red] table [x=y, y=x] {./SimulationResults/PlotData/MR_64_N_16_Stage_2_Beam_1_hbf.txt};
\addplot [no markers, thick, blue] table [x=y, y=x] {./SimulationResults/PlotData/MR_64_N_16_Stage_2_Beam_2_hbf.txt};
\end{polaraxis}
\end{tikzpicture}
}
\subfloat[][]{
\begin{tikzpicture}
\begin{polaraxis}[
	scale only axis=true,
 	width = 0.25*\columnwidth,
	height = 0.25*\columnwidth,
	grid=both,
   	rotate=-90,
	xticklabel=$\pgfmathprintnumber{\tick}^{\circ}$,
   	xtick={0, 45, 135, 180, 225, 315},
	minor xtick={90, 270},
	x dir=reverse,
   	xticklabel style={anchor=-\tick-90, font=\tiny},
	xtick style={font=\tiny},
   	ytick={-40,-30,-20,-10,0},
   	ymin=-40, ymax=5,
   	y coord trafo/.code=\pgfmathparse{#1+40},
   	y coord inv trafo/.code=\pgfmathparse{#1-40},
	yticklabels={,,},
	ymajorticks=true,
	yminorticks=true,
]
\addplot [no markers, thick, red] table [x=y, y=x] {./SimulationResults/PlotData/MR_64_N_16_Stage_3_Beam_1_hbf.txt};
\addplot [no markers, thick, blue] table [x=y, y=x] {./SimulationResults/PlotData/MR_64_N_16_Stage_3_Beam_2_hbf.txt};
\end{polaraxis}
\end{tikzpicture}
}
\caption{Beams of different width of a sub-array hybrid beamforming array.}
\label{FIG:BEAMPATTERNENQ}
\vspace*{-0.25cm}
\end{figure}
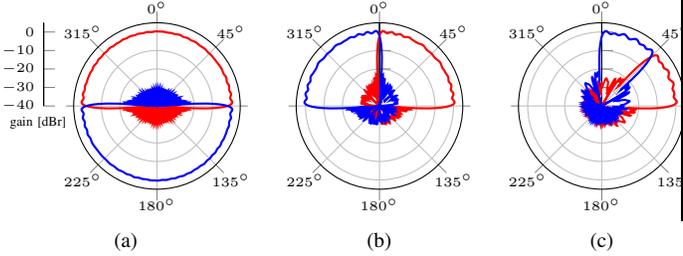
\begin{figure}
\vspace*{-0.5cm}
\centering
\subfloat[][]{
\begin{tikzpicture}
\begin{polaraxis}[
	scale only axis=true,
 	width = 0.25*\columnwidth,
	height = 0.25*\columnwidth,
  	rotate=-90,
	grid=both,
	xticklabel=$\pgfmathprintnumber{\tick}^{\circ}$,
   	xtick={0, 45, 135, 180, 225, 315},
	minor xtick={90, 270},
	x dir=reverse,
	xticklabel style={anchor=-\tick-90, font=\tiny},
	xtick style={font=\tiny},
   	ytick={-40,-30,-20,-10,0},
   	ymin=-40, ymax=5,
   	y coord trafo/.code=\pgfmathparse{#1+40},
   	y coord inv trafo/.code=\pgfmathparse{#1-40},
	ylabel style={font=\tiny, yshift=-0.11*\columnwidth, xshift=-0.20*\columnwidth},
	ylabel={gain [dBr]},
	yticklabel style={anchor=east, xshift=-0.17*\columnwidth, font=\tiny},
  	y axis line style={yshift=-0.17*\columnwidth,},
   	ytick style={yshift=-0.17*\columnwidth,font=\tiny},
]
\addplot [no markers, thick, red] table [x=y, y=x] {./SimulationResults/PlotData/MR_64_N_16_Stage_1_Beam_1_hbf_full.txt};
\addplot [no markers, thick, blue] table [x=y, y=x] {./SimulationResults/PlotData/MR_64_N_16_Stage_1_Beam_2_hbf_full.txt};
\end{polaraxis}
\end{tikzpicture}
}
\subfloat[][]{
\begin{tikzpicture}
\begin{polaraxis}[
	scale only axis=true,
 	width = 0.25*\columnwidth,
	height = 0.25*\columnwidth,
	grid=both,
   	rotate=-90,
	xticklabel=$\pgfmathprintnumber{\tick}^{\circ}$,
   	xtick={0, 45, 135, 180, 225, 315},
	minor xtick={90, 270},
	x dir=reverse,
   	xticklabel style={anchor=-\tick-90, font=\tiny},
	xtick style={font=\tiny},
   	ytick={-40,-30,-20,-10,0},
   	ymin=-40, ymax=5,
   	y coord trafo/.code=\pgfmathparse{#1+40},
   	y coord inv trafo/.code=\pgfmathparse{#1-40},
	yticklabels={,,},
	ymajorticks=true,
	yminorticks=true,
]
\addplot [no markers, thick, red] table [x=y, y=x] {./SimulationResults/PlotData/MR_64_N_16_Stage_2_Beam_1_hbf_full.txt};
\addplot [no markers, thick, blue] table [x=y, y=x] {./SimulationResults/PlotData/MR_64_N_16_Stage_2_Beam_2_hbf_full.txt};
\end{polaraxis}
\end{tikzpicture}
}
\subfloat[][]{
\begin{tikzpicture}
\begin{polaraxis}[
	scale only axis=true,
 	width = 0.25*\columnwidth,
	height = 0.25*\columnwidth,
	grid=both,
   	rotate=-90,
	xticklabel=$\pgfmathprintnumber{\tick}^{\circ}$,
   	xtick={0, 45, 135, 180, 225, 315},
	minor xtick={90, 270},
	x dir=reverse,
   	xticklabel style={anchor=-\tick-90, font=\tiny},
	xtick style={font=\tiny},
   	ytick={-40,-30,-20,-10,0},
   	ymin=-40, ymax=5,
   	y coord trafo/.code=\pgfmathparse{#1+40},
   	y coord inv trafo/.code=\pgfmathparse{#1-40},
	yticklabels={,,},
	ymajorticks=true,
	yminorticks=true,
]
\addplot [no markers, thick, red] table [x=y, y=x] {./SimulationResults/PlotData/MR_64_N_16_Stage_3_Beam_1_hbf_full.txt};
\addplot [no markers, thick, blue] table [x=y, y=x] {./SimulationResults/PlotData/MR_64_N_16_Stage_3_Beam_2_hbf_full.txt};
\end{polaraxis}
\end{tikzpicture}
}
\caption{Beams of different width of a fully-connected hybrid beamforming array.}
\label{FIG:BEAMPATTERNFNQ}
\vspace*{-0.25cm}
\end{figure}
\begin{figure}
\vspace*{-0.5cm}
\centering
\subfloat[][]{
\begin{tikzpicture}
\begin{polaraxis}[
	scale only axis=true,
 	width = 0.25*\columnwidth,
	height = 0.25*\columnwidth,
  	rotate=-90,
	grid=both,
	xticklabel=$\pgfmathprintnumber{\tick}^{\circ}$,
   	xtick={0, 45, 135, 180, 225, 315},
	minor xtick={90, 270},
	x dir=reverse,
	xticklabel style={anchor=-\tick-90, font=\tiny},
	xtick style={font=\tiny},
   	ytick={-30,-20,-10,0,10},
   	ymin=-30, ymax=10,
   	y coord trafo/.code=\pgfmathparse{#1+30},
   	y coord inv trafo/.code=\pgfmathparse{#1-30},
	ylabel style={font=\tiny, yshift=-0.11*\columnwidth, xshift=-0.20*\columnwidth},
	ylabel={gain [dB]},
	yticklabel style={anchor=east, xshift=-0.17*\columnwidth, font=\tiny},
  	y axis line style={yshift=-0.17*\columnwidth,},
   	ytick style={yshift=-0.17*\columnwidth,font=\tiny},
]
\addplot [no markers, thick, red] table [x expr=\thisrow{y}+90, y=x] {./SimulationResults/PlotData/MR_64_N_16_Stage_1_Beam_1_hbf_full_quant_2bit.txt};
\addplot [no markers, thick, blue] table [x expr=\thisrow{y}+90, y=x] {./SimulationResults/PlotData/MR_64_N_16_Stage_1_Beam_2_hbf_full_quant_2bit.txt};
\end{polaraxis}
\end{tikzpicture}
}
\subfloat[][]{
\begin{tikzpicture}
\begin{polaraxis}[
	scale only axis=true,
 	width = 0.25*\columnwidth,
	height = 0.25*\columnwidth,
	grid=both,
   	rotate=-90,
	xticklabel=$\pgfmathprintnumber{\tick}^{\circ}$,
   	xtick={0, 45, 135, 180, 225, 315},
	minor xtick={90, 270},
	x dir=reverse,
   	xticklabel style={anchor=-\tick-90, font=\tiny},
	xtick style={font=\tiny},
   	ytick={-30,-20,-10,0,10},
   	ymin=-30, ymax=10,
   	y coord trafo/.code=\pgfmathparse{#1+30},
   	y coord inv trafo/.code=\pgfmathparse{#1-30},
	yticklabels={,,},
	ymajorticks=true,
	yminorticks=true,
]
\addplot [no markers, thick, red] table [x expr=\thisrow{y}+90, y=x] {./SimulationResults/PlotData/MR_64_N_16_Stage_2_Beam_2_hbf_full_quant_2bit.txt};
\addplot [no markers, thick, blue] table [x expr=\thisrow{y}+90, y=x] {./SimulationResults/PlotData/MR_64_N_16_Stage_2_Beam_1_hbf_full_quant_2bit.txt};
\end{polaraxis}
\end{tikzpicture}
}
\subfloat[][]{
\begin{tikzpicture}
\begin{polaraxis}[
	scale only axis=true,
 	width = 0.25*\columnwidth,
	height = 0.25*\columnwidth,
	grid=both,
   	rotate=-90,
	xticklabel=$\pgfmathprintnumber{\tick}^{\circ}$,
   	xtick={0, 45, 135, 180, 225, 315},
	minor xtick={90, 270},
	x dir=reverse,
   	xticklabel style={anchor=-\tick-90, font=\tiny},
	xtick style={font=\tiny},
   	ytick={-30,-20,-10,0,10},
   	ymin=-30, ymax=10,
   	y coord trafo/.code=\pgfmathparse{#1+30},
   	y coord inv trafo/.code=\pgfmathparse{#1-30},
	yticklabels={,,},
	ymajorticks=true,
	yminorticks=true,
]
\addplot [no markers, thick, red] table [x expr=\thisrow{y}+180, y=x] {./SimulationResults/PlotData/MR_64_N_16_Stage_3_Beam_2_hbf_full_quant_2bit.txt};
\addplot [no markers, thick, blue] table [x expr=\thisrow{y}+180, y=x] {./SimulationResults/PlotData/MR_64_N_16_Stage_3_Beam_1_hbf_full_quant_2bit.txt};
\end{polaraxis}
\end{tikzpicture}
}
\caption{Beams of different width optimized for sidelobe attenuation and with 2 bit quantization of the phase shifters of a fully-connected hybrid beamforming.}
\label{FIG:BEAMPATTERNFQ}
\vspace*{-0.25cm}
\end{figure}
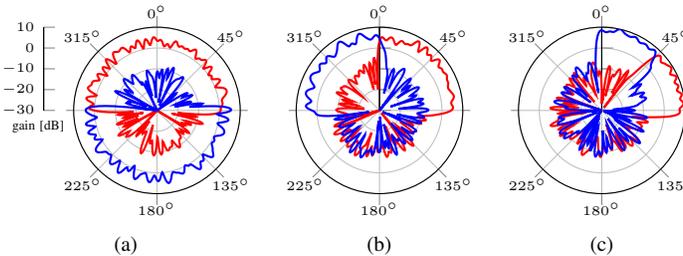
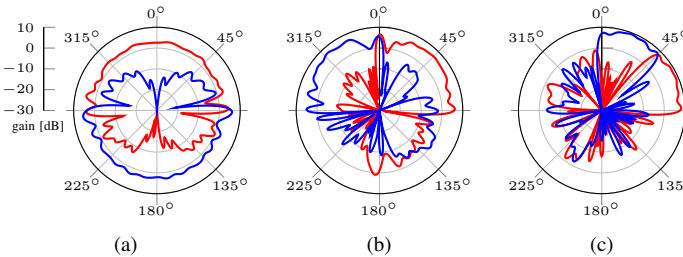
\begin{figure}
\vspace*{-0.5cm}
\centering
\subfloat[][]{
\begin{tikzpicture}
\begin{polaraxis}[
	scale only axis=true,
 	width = 0.25*\columnwidth,
	height = 0.25*\columnwidth,
  	rotate=-90,
	grid=both,
	xticklabel=$\pgfmathprintnumber{\tick}^{\circ}$,
   	xtick={0, 45, 135, 180, 225, 315},
	minor xtick={90, 270},
	x dir=reverse,
	xticklabel style={anchor=-\tick-90, font=\tiny},
	xtick style={font=\tiny},
   	ytick={-30,-20,-10,0,10},
   	ymin=-30, ymax=10,
   	y coord trafo/.code=\pgfmathparse{#1+30},
   	y coord inv trafo/.code=\pgfmathparse{#1-30},
	ylabel style={font=\tiny, yshift=-0.11*\columnwidth, xshift=-0.20*\columnwidth},
	ylabel={gain [dB]},
	yticklabel style={anchor=east, xshift=-0.17*\columnwidth, font=\tiny},
  	y axis line style={yshift=-0.17*\columnwidth,},
   	ytick style={yshift=-0.17*\columnwidth,font=\tiny},
]
\addplot [no markers, thick, red] table [x expr=\thisrow{y}+90, y=x] {./SimulationResults/PlotData/MR_64_N_16_Stage_1_Beam_1_hbf_full_imdea.txt};
\addplot [no markers, thick, blue] table [x expr=\thisrow{y}+90, y=x] {./SimulationResults/PlotData/MR_64_N_16_Stage_1_Beam_2_hbf_full_imdea.txt};
\end{polaraxis}
\end{tikzpicture}
}
\subfloat[][]{
\begin{tikzpicture}
\begin{polaraxis}[
	scale only axis=true,
 	width = 0.25*\columnwidth,
	height = 0.25*\columnwidth,
	grid=both,
   	rotate=-90,
	xticklabel=$\pgfmathprintnumber{\tick}^{\circ}$,
   	xtick={0, 45, 135, 180, 225, 315},
	minor xtick={90, 270},
	x dir=reverse,
   	xticklabel style={anchor=-\tick-90, font=\tiny},
	xtick style={font=\tiny},
   	ytick={-30,-20,-10,0,10},
   	ymin=-30, ymax=10,
   	y coord trafo/.code=\pgfmathparse{#1+30},
   	y coord inv trafo/.code=\pgfmathparse{#1-30},
	yticklabels={,,},
	ymajorticks=true,
	yminorticks=true,
]
\addplot [no markers, thick, red] table [x expr=\thisrow{y}+90, y=x] {./SimulationResults/PlotData/MR_64_N_16_Stage_2_Beam_1_hbf_full_imdea.txt};
\addplot [no markers, thick, blue] table [x expr=\thisrow{y}+90, y=x] {./SimulationResults/PlotData/MR_64_N_16_Stage_2_Beam_2_hbf_full_imdea.txt};
\end{polaraxis}
\end{tikzpicture}
}
\subfloat[][]{
\begin{tikzpicture}
\begin{polaraxis}[
	scale only axis=true,
 	width = 0.25*\columnwidth,
	height = 0.25*\columnwidth,
	grid=both,
   	rotate=-90,
	xticklabel=$\pgfmathprintnumber{\tick}^{\circ}$,
   	xtick={0, 45, 135, 180, 225, 315},
	minor xtick={90, 270},
	x dir=reverse,
   	xticklabel style={anchor=-\tick-90, font=\tiny},
	xtick style={font=\tiny},
   	ytick={-30,-20,-10, 0, 10},
   	ymin=-30, ymax=10,
   	y coord trafo/.code=\pgfmathparse{#1+30},
   	y coord inv trafo/.code=\pgfmathparse{#1-30},
	yticklabels={,,},
	ymajorticks=true,
	yminorticks=true,
]
\addplot [no markers, thick, red] table [x expr=\thisrow{y}+90, y=x] {./SimulationResults/PlotData/MR_64_N_16_Stage_3_Beam_1_hbf_full_imdea.txt};
\addplot [no markers, thick, blue] table [x expr=\thisrow{y}+90, y=x] {./SimulationResults/PlotData/MR_64_N_16_Stage_3_Beam_2_hbf_full_imdea.txt};
\end{polaraxis}
\end{tikzpicture}
}
\caption{Beams of different width  of fully-connected hybrid beamforming array with phase quantization according to \cite{Palacios2016}.}
\label{FIG:BEAMPATTERNFQIMDEA}
\vspace*{-0.25cm}
\end{figure}
\begin{table}
	\renewcommand{\arraystretch}{1.3}
	\caption{Comparison of the designed beams.}
	\label{TAB:measuretab}
	\centering
		\begin{tabular}{|p{2.565cm}|p{1cm}|p{1cm}|p{1cm}|p{1.1cm}|}
			\hline 
			Beam & avg. gain dB & max ripple dB & overlap in \% & max side-lope dB\\ \hline \hline
			Fig. \ref{FIG:BEAMPATTERNENQ} (a) & 18.2 & 4.00 & 2.44 & -17.4 \\ \hline
			Fig. \ref{FIG:BEAMPATTERNENQ} (b) & 21.7 & 2.89 & 3.22 & -16.2 \\ \hline
			Fig. \ref{FIG:BEAMPATTERNENQ} (c) & 26.3 & 2.76 & 7.21 & -16.3 \\ \hline
			Fig. \ref{FIG:BEAMPATTERNFNQ} (a) & 18.2 & 2.04 & 2.63 & -22.6 \\ \hline
			Fig. \ref{FIG:BEAMPATTERNFNQ} (b) & 22. 0 & 2.10 & 2.63 & -22.8 \\ \hline
			Fig. \ref{FIG:BEAMPATTERNFNQ} (c) & 24.8 & 2.35 & 5.26 & -23.3 \\ \hline
			Fig. \ref{FIG:BEAMPATTERNFQ} (a) & 2.52 & 3.90 & 7.66 & -10.3 \\ \hline
			Fig. \ref{FIG:BEAMPATTERNFQ} (b) & 5.50 & 3.01 & 6.54 & -10.1 \\ \hline
			Fig. \ref{FIG:BEAMPATTERNFQ} (c) & 8.23 & 1.47 & 6.63 & -12.7 \\ \hline
			Fig. \ref{FIG:BEAMPATTERNFQIMDEA} (a) & 2.22 & 8.82 & 34.4 & -2.16 \\ \hline
			Fig. \ref{FIG:BEAMPATTERNFQIMDEA} (b) & 5.04 & 7.25 & 8.20 & -4.04 \\ \hline
			Fig. \ref{FIG:BEAMPATTERNFQIMDEA} (c) & 8.02 & 1.49 & 14.4 & -8.97 \\ \hline
		\end{tabular}
\end{table}
\section{Conclusion}
The developed approach can synthesize any beam-pattern for hybrid-beamforming systems. The numerical examples showed that
a sufficient solution to the underlying optimization problem can be found with reasonable computational complexity. The numeric examples also demonstrated that
it is possible to adapt the approach to any type of constraint arising in the context of hybrid beamforming and wireless communication. 

If we compare the beams synthesized with the method introduced in this method to the ones in \cite{Palacios2016} we can achieve a significant smaller overlap 
7.66 \%, 6.54 \% and 6.63 \% compared to 34.4 \$, 8.20 \%, and 14.4 \%. This beams are designed for a hierarchical beam search, thus the max side lope is a especially 
important criteria. Here our result of -10.3 dB, -10.1 dB and -12.7 dB is also significantly better than -2.16 dB, -4.04 dB and -8.79 dB.

\section*{Acknowledgment}
The research leading to these results received funding from the European Commission H2020 programme under grant agreement no 671650 (5G PPP mmMAGIC project).

\bibliographystyle{IEEEtran}
\bibliography{./literature/IEEEabrv,../../../bibliography/bibKilian}

\end{document}